\documentclass[12pt]{article}
\usepackage{graphicx}

\newcommand\pubnumber{DPF2015-238}
\newcommand\pubdate{\today}

\def\napoli{Department of Physics and Astronomy,\\
Seoul National University, Seoul, Korea}

\def\Title#1{\begin{center} {\Large #1 } \end{center}}
\def\Author#1{\begin{center}{ \sc #1} \end{center}}
\def\Address#1{\begin{center}{ \it #1} \end{center}}

\newcommand\pubblock{\rightline{\begin{tabular}{l} \pubnumber\\
         \pubdate  \end{tabular}}}
\newenvironment{Abstract}{\begin{quotation}  }{\end{quotation}}
\newenvironment{Presented}{\begin{quotation} \begin{center} 
             PRESENTED AT\end{center}\bigskip 
      \begin{center}\begin{large}}{\end{large}\end{center} \end{quotation}}
\def\Acknowledgments{\bigskip  \bigskip \begin{center} \begin{large}
             \bf ACKNOWLEDGMENTS \end{large}\end{center}}

\def\beq{\begin{equation}}
\def\eeq#1{\label{#1}\end{equation}}
\def\eeqn{\end{equation}}

\def\beqa{\begin{eqnarray}}
\def\eeqa#1{\label{#1}\end{eqnarray}}
\def\eeqan{\end{eqnarray}}

\let\bar=\overbar

\def\Dslash{\not{\hbox{\kern-4pt $D$}}}
\def\dslash{\not{\hbox{\kern-2pt $\del$}}}

\def\msb{{\bar{\ssstyle M \kern -1pt S}}}

\makeatletter
\newcommand*{\rom}[1]{\expandafter\@slowromancap\romannumeral #1@}
\makeatother

\begin{document}
\begin{titlepage}
\pubblock

\vfill
\Title{Status of the KIMS-NaI experiment}
\vfill
\Author{Kyungwon Kim on Behalf of the KIMS Collaboration}
\Address{\napoli}
\vfill
\begin{Abstract}
KIMS-NaI is a direct detection experiment searching for Weakly Interacting Massive Particles (WIMP) via their scattering off of nuclei in a NaI(Tl) crystal.
The KIMS-NaI collaboration has carried out tests of six crystals in the Yangyang underground laboratory in order to develope low-background NaI(Tl) crystals.
Studies of internal backgrounds crystals have been performed with the goal of reducing backgrounds levels to 1 dru at 2 keV.
Pulse shape discrimination (PSD) capabilities were also investigated for distinguishing between WIMP nuclear recoil signals and electron recoil backgrounds.
The PSD analysis was applied to underground data with one low background NaI(Tl) detector and the evaluation of WIMP mass limit is ongoing.
\end{Abstract}
\vfill
\begin{Presented}
DPF 2015\\
The Meeting of the American Physical Society\\
Division of Particles and Fields\\
Ann Arbor, Michigan, August 4--8, 2015\\
\end{Presented}
\vfill
\end{titlepage}
\def\thefootnote{\fnsymbol{footnote}}
\setcounter{footnote}{0}

\section{Introduction}
Numerous observations provide evidence that the existence of non-baryonic dark matter is a dominent component of the Universe~\cite{dm1,dm2}.
WIMPs are well-supported dark matter candidates in many theories~\cite{wimp}.
In recent years, there is unresolved conflict in low mass WIMP region.
Among many direct detecion experiments, DAMA/LIBRA claimed the observation of an annual modulation signature of WIMPs~\cite{DAMA2} while other experiments, such as XENON100~\cite{XENON}, LUX~\cite{LUX} and SuperCDMS~\cite{SCDMS}, reported null signals.
The KIMS-NaI experiment aims to confirm the DAMA/LIBRA observations using the same NaI(Tl) scintillating crystal target.

Since the interaction rate of dark matter WIMPs is very low, the suppresion of background events is an essential requirement for dark matter search experiment.
We have tested six crystals and studied their internal background levels as part of a program to develope ultra-low background crystal.
Another rejection techinique used in the KIMS experiment is the discrimination of WIMP-nuclear recoil signals from electron recoils using PSD.
This article describes the background level measurement results and a PSD analysis study of newly developed NaI(Tl) crystal. We also discuss and future prospects.

\section{KIMS-NaI detector}

KIMS detectors were installed inside a substantial shield to achieve a low background crystal from external radiactivity and cosmic rays in the Yangyang underground laboratory.
NaI(Tl) detectors were placed between the CsI(Tl) detector array used for the KIMS-CsI experiment~\cite{kims_plb,kims_prl,kims_prl2}.
Six cylindrical NaI(Tl) crystals were used in the test.
Among six crystals, five crystals were produced by Alpha Spectra (AS) and one by Beijing Hamamatus (BH).
NaI-001 and NaI-002 were grown from AS powder and produced by them.
The test result of these crystals is described in ~\cite{kims_nai}.
NaI-003 grown from Astro Grade (AG) powder was produced by Sigma Aldrich (SA) which is known to less contamination from $^{40}$K.
NaI-004 and NaI-006 were grown from Crystal Grade (CG) which is a commercial grade powder and produced by different company AS and BH.
NaI-005 was grown from special powder aimed at reducing $^{210}$Pb.
Scintillation light signals from crystals are read out by high quantum efficiency PMTs, R12669 made by Hamamatsu company.
The specifications of six crystals are listed in Table \ref{tab_crystal}.

\begin{table}[t]
\begin{center}
\begin{tabular}{l|ccccccc}
 		&  Mass (kg)	& D $\times$ L			& Powder	& Crystal		& Arrival	& LY (PE/keV)	\\ \hline
 NaI-001	& 8.26		& 5$''$ $\times$ 7$''$		& AS		& AS			& 2013.9	& 15.60$\pm$1.41	\\
 NaI-002	& 9.15		& 4.2$''$ $\times$ 11$''$	& AS		& AS 		& 2014.1	& 15.51$\pm$1.41	\\
 NaI-003	& 3.3		& 4.5$''$ $\times$ 3.5$''$	& SA-AG	& AS			& 2014.8	& 13.26$\pm$1.28	\\
 NaI-004	& 3.3		& 4.5$''$ $\times$ 3.5$''$	& SA-CG	& AS			& 2014.8	& 3.85$\pm$0.38	\\
 NaI-005	& 9.16		& 4.5$''$ $\times$ 11$''$	& AS-WS\rom{2}	& AS	& 2014.12	& 12.14$\pm$1.14	\\
 NaI-006	& 11.44		& 110 $mm$ $\times$ 200 $mm$	& SA-CG	& BH			& 2015.1	& 4.36$\pm$0.39	\\ \hline
\end{tabular}
\caption{Specifications of KIMS-NaI Crystals}
\label{tab_crystal}
\end{center}
\end{table}

\section{Event Selections}

A selection of scintillation events is required to remove unwanted background events.
Firstly, events having less than two photoelectrons are discarded in any of PMTs to distinguish scintillation events from PMT noise. 

Parameters related to charge ratio which were developed by the DAMA/LIBRA~\cite{dama_x1x2} and charge asymmetry were used for event selections.
The variables for the ratio of the pulse areas exploits different time characteristics of fast noise pulses and relatively slow scintillation pulses.
Fast and slow charge ratio are denoted by $"$X1$"$ and $"$X2$"$ respectively defined as follows.

\begin{equation}
X1 = \frac{Q_{100~to~600 ns}}{Q_{0~to~600 ns}},
X2 = \frac{Q_{0~to~50 ns}}{Q_{0~to~600 ns}}
\end{equation}
where Q is charge in specific time range.

Charge asymmetric events produced by different sharing of the light colliection depending on the position of the interaction inside crystal are observed in the background data.
We defined the charge asymmetry varible as 
\begin{equation}
Asymmetry = \frac{Q_{1}-Q_{2}}{Q_{1}+Q_{2}}
\label{eq_asym}
\end{equation}
where Q$_{1}$ and Q$_{2}$ are charge from each side PMT.

\begin{figure}[htb]
\centering
\includegraphics[height=2.5in]{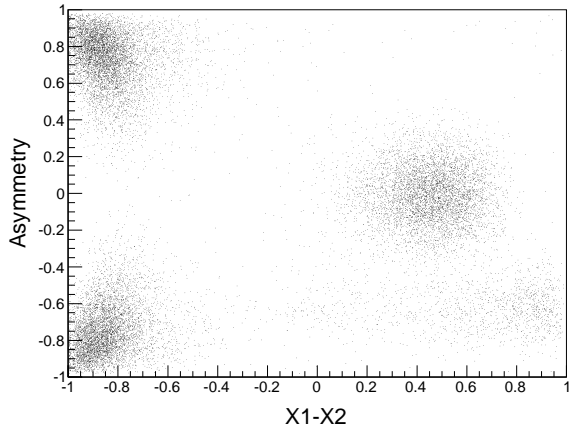}
\caption{Two-dimensional asymmetry and X1-X2 for background in 4-5 keV energy region.}
\label{fig_xas}
\end{figure}

Figure \ref{fig_xas} shows that two-dimensional asymmetry and X1-X2 for underground data in 4-5 keV energy region.
We identify scintillation signals which have large X1 and small X2 values based on gamma source data, while noise signals generally have small X1 and large X2 values~\cite{kims_nai}.
Large values of X1-X2 events and near zero asymmetry events are selected as a WIMP candidate event.

\section{Backgrounds of the KIMS-NaI}

The most relevant contribution to background are coming from natural radioisotopes in the crystal.
Measured values of the contamination levels in the six crystals are listed in Table \ref{tab_bkg}.

We estimated contamination from $^{40}$K which is dominant background especially in 3 keV energy region.
The $^{40}$K decay can be estimated by searching coincdence signals of 3 keV deposition in the NaI(Tl) and the 1460 keV $\gamma$ escaping from it and hit in CsI(Tl) detectors which surround NaI(Tl) detector.
As shown in Table \ref{tab_bkg}, NaI-003 grown from SA-AG powder has $^{40}$K levels about 25 ppb which is the lowest level of $^{40}$K contamination.
It shows that the source of $^{40}$K contamination in the NaI(Tl) crystals originates mostly from the NaI powder.
Several approaches are currently under development for reduction of $^{40}$K contamination.

The contaminations from $^{238}$U and $^{232}$Th are quantified by evaluating time interval between $\alpha$-$\beta$ events and $\alpha$-$\alpha$ events respectively after identifying $\alpha$ events taking advantage of pulse shape differneces.
Then, by fitting the obtained time interval and deriving the activities of radioisotopes in each chain, the contamination level can be estimated.
The $^{238}$U background level is less than 0.0015 mBq/kg and The $^{232}$Th background level is 0.002 $\pm$ 0.001 mBq/kg in the NaI-002.
The measured contamination level of $^{238}$U and $^{232}$Th are small.

Total alpha rate of the NaI-002 is 1.77 $\pm$ 0.01 mBq/kg which is the large value compare to the contamination level of $^{238}$U and $^{232}$Th.
This suggests that they are due to decays of $^{210}$Po nuclei originating from $^{222}$Rn contamination during the powder or crystal processing.
We estimated crystal manufactured date using total alpha rate change and this evaluation coincides with the time that the crystal was grown.
A crystal growing incorporating with crystal companies is on-going to reduce contamination level of $^{210}$Pb.
NaI-005 grown from AS-WS\rom{2} experiencing better treatment on powder process has the lowest contaimnation level of $^{210}$Pb.
We are also investigaing powder purification test with various resin and measurement of alpha in the powder.

Internal backgrounds level will be reduced to 10 ppb of $^{40}$K and 0.2 mBq/kg of $^{210}$Pb via a combination of these reduction techniques.
External backgrounds and some remained internal backgrounds can be vetoed by liquid scintillator veto system with more than 80\% efficiency based on the GEANT4 simulation.
These implementations will give 1 dru level of background to KIMS-NaI detectors.

Details about background understanding of six NaI(Tl) crystals is described in ~\cite{kims_naibkg}

\begin{table}[t]
\begin{center}
\begin{tabular}{l|ccc}
 		& K-Crystal		& K-Powder	& Total $\alpha$	\\
 		& (ppb)			& (ppb)		& (mBq/kg)		\\ \hline
 NaI-001	& 41.4$\pm$2.99	&	-		& 3.29$\pm$0.01	\\
 NaI-002	& 49.3$\pm$2.43	&	-		& 1.77$\pm$0.01	\\
 NaI-003	& 25.43$\pm$3.57	& 25.07		& 2.43$\pm$0.011	\\
 NaI-004	& 116.70$\pm$6.78	& $\sim$ 200	&	-			\\
 NaI-005	& 40.1$\pm$4.2	&	-		& 0.48$\pm$0.004	\\
 NaI-006	& 127.12$\pm$6.45	& $\sim$ 200	& 1.53$\pm$0.007	\\ \hline
\end{tabular}
\caption{Measurements of internal backgrounds in the NaI(Tl) crystals. First and second columns are measured level of K in the crystal and the powder respectively.}
\label{tab_bkg}
\end{center}
\end{table}

\section{Pulse Shape Discrimination Analysis}

The different time distributions of the photoelectrons produced by nuclear and electron recoils in the scintillation crystal make discriminate WIMP-nuclear recoil signals from electron recoil backgrounds.
PSD parameter to characterize the time distribution is defined as
\begin{equation}
ln(MT) = ln \Big(\frac{\sum A_{i}t_{i}}{\sum A_{i}}-t_{0}\Big),
\label{eq:meantime}
\end{equation}
where $A_{i}$ is charge of the ith cluster, $t_{i}$ is the time of the ith cluster and $t_{0}$ is the time of the first cluster.

\begin{figure}[htb]
\centering
\includegraphics[height=2.2in]{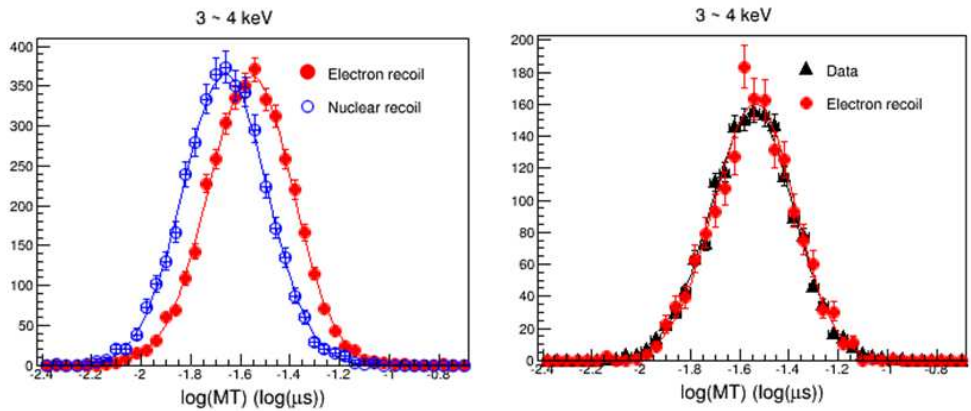}
\caption{ln(MT) distribution of electron recoil and nuclear recoil sample (left), electron recoil and background data (right)}
\label{fig_psd_all}
\end{figure}

A small crystal from same ingot with the NaI-002 was exposed to Am/Be source and $^{137}$Cs for nuclear recoil and electron recoil measurement respectively.
PSD measurements of newly developed NaI crystal shows that it has better PSD power than previous measurements~\cite{psd_other} and our results using CsI crystal due to high light output of the crystal.
The ln(MT) distribution measured in small crystal is in the left figure in Figure \ref{fig_psd_all} which shows well discrimination of nuclear recoil form electron recoil.
Experimental setup and measurment result are described in ~\cite{kims_csipsd, kims_naipsd} in detail.

133.7 days of underground data with the NaI-002 was analyzed using PSD.
For underground data, multiple hit events where more than one detectors are hit were considered as electron recoil samples and single hit events where only one of detectors is hit is used for the analysis.
The right figure in Figure \ref{fig_psd_all} shows that ln(MT) distribution of backgrounds and electron recoil in underground data.
The extraction of nuclear recoil rate from data and setting limit of WIMP mass using this data is in progress.

\section{Prospects of the KIMS-NaI}

\begin{figure}[htb]
\centering
\includegraphics[height=3.0in]{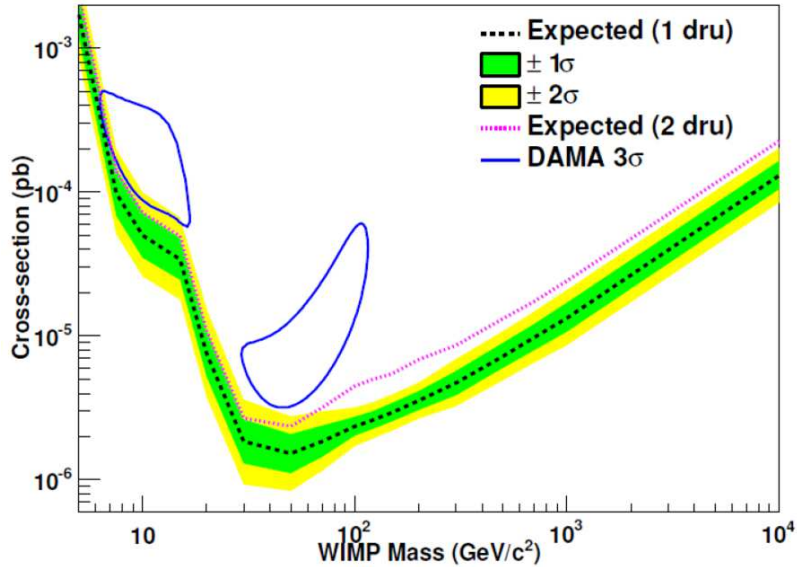}
\caption{Expected sensitivity with annual modulation analysis using 200 kg, 3 years of data}
\label{fig_limit}
\end{figure}

We plan the experiment with 200 kg of NaI(Tl) crystal after reaching a 1 dru background level.
To achieve a 1 dru background level, internal backgrounds and external backgrounds sholud be reduced to 0.5 dru repectively.
By developing ultra-pure crystals, internal backgrounds can be reduced to more than a factor of two.
The liquid scintillation veto system to reduce external backgrounds will be deployed within active shielding and the prototype is currently operating.
Figure \ref{fig_limit} shows that expected sensitivity of WIMP-nucleon spin independent interactions using the annual modulation analysis with total mass of 200 kg and 3 years of operation under 2 keV energy threshold assuming a standard halo model.
We expect that the annual modulation study with this 3 years of data make investigate the DAMA/LIBRA 3$\sigma$ region as shown in the figure.

\section{Conclusions}
The KIMS-NaI aims at directly detecting of WIMP dark matter with NaI(Tl) crystals.
The collaboration is currently working on R$\&$D to develope ultra-low background crystals.
The operation of 200 kg with 1 dru background NaI(Tl) crystals will enable a test of the DAMA/LIBRA annual modulation signature.
The PSD analysis is investigating with the crystal having high light output and low backgrounds.
This will enhance the credibility of the annual modulation study.

\Acknowledgments
This work was funded by the Institute for Basic Science (IBS) under project code IBS-R016-D1 and was supported by the Basic Science Research Program through the National Research Foundation of Korea funded by the Ministry of Education (NRF-2011-35B-C00007).

\end{document}